# Open-source automated chemical vapor deposition system for the production of two- dimensional nanomaterials

Lizandra Williams- Godwin[1,¶], Dale Brown[1,¶], Richard Livingston[2], Tyler Webb[1], Lynn Karriem[1], Elton Graugnard[1], David Estrada[1,3*]

[1]Micron School of Materials Science & Engineering, Boise State University, Boise, Idaho, United States of America

[2]Department of Mechanical and Biomedical Engineering, Boise State University, Boise, Idaho, United States of America

[3]Center for Advanced Energy Studies, Boise State University, Boise, Idaho, United States of America

\* Corresponding author

Email: daveestrada@boisestate.edu (DE)

¶These authors contributed equally to this work.

# Abstract

The study of two- dimensional (2D) materials is a rapidly growing area within nanomaterial research. However, the high equipment costs, which include the processing systems necessary for creating these materials, can be a barrier to entry for some researchers interested in studying these novel materials. Such process systems include those used for chemical vapor deposition. This article presents the first open-source design for an automated chemical vapor deposition system that can be built for less than a third of the cost for a similar commercial system. Our design can be easily customized and expanded on, depending upon the needs of the user. With a process chamber built as described, we demonstrate that a variety of 2D nanomaterials and their heterostructures can be grown via chemical vapor deposition. Specifically, our experimental results demonstrate the capability of this open-source design in producing high quality, 2D nanomaterials such as graphene and tungsten disulfide, which are at the forefront of research in emerging semiconductor devices, sensors, and energy storage applications.

# Introduction

The modern research push into two- dimensional (2D) materials was sparked by Novoselov and Geim's isolation of few- and single-layer graphene via the mechanical exfoliation of graphite [1]. Novoselov et al. discovered that these single- and few-atom thick crystals of graphitic carbon, known as graphene, were stable in air, semi-metallic, and had extremely high carrier mobility (about 10,000 $cm^2$ / V-s), though the on/off ratio of graphene transistors was found to be less than 30 at room temperature, limiting its usefulness as an active component in transistors for digital CMOS circuits. Since then, 2D materials such as hexagonal boron nitride

(hBN) [2] and tungsten disulfide (WS$_2$) [3] among a host of others [4-10] are garnering much research attention due to their potential use in emerging semiconductor device applications. Moreover, chemical vapor deposition (CVD) is one of the most common methods for the production of these novel 2D materials [3], [9], [11-14].

In the 2D materials and nanotechnology research community, there is a common barrier that potentially slows progress in the field. This barrier is due to the high cost of systems such as the CVD chamber that is needed in order to synthesize high quality films of these advanced materials. CVD is a synthesis process widely used in the production of commercial microelectronics, but the equipment for this process is very costly to purchase and often too expensive for the average researcher.

A viable solution to this challenge is open-source technology. Open-source technology, broadly speaking, is the field of technological development in which the design of the final product is published and freely available for others to use and adapt to their own needs. This is in contrast to the standard commercial approach to hardware development where designs and software are often patented or kept as trade secrets, giving the company an advantage in the development and use of their technology [15]. Open-source hardware, such as the RepRap – an open-source 3D printer [16], syringe pumps [17], optics equipment [18] and many others [19-21], are freely available to the user community and has led to even further accessibility in research. One of the chief benefits of open-source technology for scientific research lies in the significant reduction of the engineering cost and time to conceptualize and develop the hardware. A second benefit of an open-source design is the community of users linked through a common experience to discuss possible methods of use and troubleshooting ideas. A similar argument can be made for the software required to automate these systems.

This paper describes a feasible option for those that require a CVD system for their materials research but also have financial limitations that prevent the purchase of a commercial system. We have designed, assembled, and programmed an automated CVD system capable of synthesizing high-quality 2D materials for about $30,000 USD in hardware costs. This paper provides the details of the design and the software making it an open-source system to the research community. Researchers adapting this design will be able to customize the system to their needs and have confidence that the system is capable of producing high-quality 2D materials.

## Materials and methods

Using the design described in this manuscript, a four gas, variable pressure CVD system, as shown in Fig 1, can be built for around $30,000 USD. A three gas, atmospheric pressure CVD system could be built for approximately $16,000 USD worth of parts. The bill of materials is contained in the supporting information (S1 Appendix) and is divided by subsystem such that the builder can easily swap out parts within a subsystem, add or remove entire subsystems, or even adapt subsystems and their respective LabVIEW drivers for use in other systems. The intake side (Fig 1A) includes the following subsystems: (1) gas delivery lines, (2) gas filter lines, (3) mass flow controllers, (4) gas mixing line, (5) intake manifold, and (6) pressure gauges. The deposition chamber (Fig 1B) is comprised of a quartz reaction tube and a Lindberg Blue M tube furnace. The exhaust side (Fig 1C) is divided into the following subsystems: (1) exhaust manifold, (2) overpressure relief, (3) vent valve, (4) exhaust line, (5) butterfly valve, (6) vacuum pump, and (7) snorkel manifold.

**Fig 1. Photograph of the variable pressure CVD system with 4 gas line option.**
The figure is annotated to indicate the (A) intake side, (B) deposition chamber, and (C) exhaust side of the system.

A brief overview of the main subsystems is given in this section. To assist with the system build, detailed construction notes are included in the supporting information, S2 Appendix. The construction notes are inclusive of all of the details needed for building the system including the order of assembly, high resolution subsystem images clearly labeled with part numbers, cable pinouts, control program setup, and user instructions for the automated program.

## Gas and pressure monitor system

The gas canisters for the system were placed into a gas cabinet immediately behind the table holding the CVD machine. If a different location is chosen for the gas cabinet, this subsystem may require some modifications. Individual gas filters are placed immediately following the gas delivery lines as shown in Fig 2. The mass flow controllers (MFCs) chosen were the M100 series from MKS Instruments (comparable MKS MFCs available today would be the G-Series). Flow rates were chosen to be 1000 sccm for argon and methane, 500 sccm for hydrogen. A 20 sccm MFC was installed for future expansion. The gases exiting the mass flow controllers are mixed and then routed through a manual diaphragm valve, which is connected to the intake manifold. The diaphragm valve is useful for system seal check and leak isolation as it isolates the MFCs and everything upstream from the remainder of the system. For variable pressure CVD applications, two capacitive manometers were used. One with a 1000 Torr full range, and the other with a 10 Torr full range. These manometers are connected to the intake manifold but can be moved to other KF25 fittings throughout the system to aid in leak detection. Both pressure gauge readings were fed into an integrated butterfly valve controller, which uses the downstream vacuum to match the system pressure to a software- controlled set point.

**Fig 2. Side view of the intake side of CVD system.**
The photograph shows the (A) gas lines, (B) gas filter lines, (C) MFCs, and (D) intake manifold.

## Reaction chamber

The furnace is a Lindberg Blue M Mini-Mite Tube Furnace. This furnace requires a 25.4mm outer diameter tube and can operate at temperatures up to 1100 ºC. Control of the furnace is possible with the controller on the front of the furnace, or via a serial connection. The reaction tube is a 609.6 mm long, 25.4 mm outer diameter sintered quartz tube that is coupled to the intake manifold via a KF25 to 25.4 mm quick disconnect and the exhaust manifold via a 25.4 mm quick disconnect to a KF40, as shown in Fig 3. The discoloration seen in the reaction tube downstream of the furnace center (Fig 3B) is mainly due to the copper deposition from graphene growths. Quartz tubes with fire-polished ends should be used to lower the likelihood of tube breakage during handling. The reaction tube can be easily changed by loosening the quick disconnects (Figs. 3A and 3C).

**Fig 3. CVD system opened to show reaction tube assembly.**
The quick connects (A) and (C) are identified, as well as the reaction tube (B).

## Data acquisition and system control

While communication with the furnace and butterfly valve was carried out via serial connections, the MFCs and capacitance gauges required analog inputs and outputs (digital communications options are available for purchase). These analog signals were supplied via a National Instruments CompactDAQ data acquisition system. The final design uses one 16-channel analog input module and two 4-channel analog output modules. Once a sample is placed into the reaction tube, the manual gas line and gas mixing valve are turned on, the furnace is turned on, and the vacuum pump is started, while all system controls are conducted on a computer running LabVIEW. The main elements of the controls programming are called virtual instruments, or VIs. For our system, the VIs include the device drivers, the manual control program, and the automatic control program. All of the VIs used to control our system are included in the supporting files (S1 Folder Programs). Detailed system startup instructions using the VI are given in the construction notes (S1 Appendix).

The device drivers are small LabVIEW programs that operate the system's active hardware (MFCs, pressure gauges, furnace, and butterfly valve). These driver programs can be opened and tested individually in order to test the operation of single active components. Or, in the case that the system design is modified, the drivers for the modified components can be reprogrammed independently, making it simple to customize the system design in order to fit individual needs.

The manual program (ManualControls.vi) controls the LabVIEW user interface. A screen capture of the graphical user interface for the manual control program is shown in Fig 4. Once the system has been built and the active components have been tested, the manual controls VI can be used to check system seal under vacuum and run simple growths. The left side of the interface is where a user can input the manual growth conditions such as pressure set point, gas flow rate, and furnace temperature. This interface provides feedback on the system conditions in real time on the right side of the interface. The first row of plots indicates the system pressure and conveniently lights up as each pressure threshold is met. The second row provides real time information on the gas flow for each MFC. Again, the buttons below the plots light up once the condition has been met. When the manual control program is in use, both setpoints and actual values for the system pressure, furnace temperature, and gas flow rates are recorded in a dated log file. The interface is set up to be easily controlled and interpreted. To gain even greater accessibility, the open-source programming language Python [22] can be used to interface with the National Instruments data acquisition hardware using the PyDAQmx package [23].

**Fig 4. Screen capture of the CVD system manual control graphical user interface.**

For complex recipes and repeatable growths, the automated control program (Read&RunRecipe.vi) should be used. The automatic control program operates the CVD system through user-generated recipes. A sample recipe *'Quick.csv'*, is included in the supporting information. The contents of the comma separated value (csv) file is shown in Table 1. Users can use this recipe as a guide to create custom recipes for their desired process. Recipes are organized into stages, where Stage 0 sets the initial setpoints and then immediately ends. The next stages, starting at Stage 1, contain the process steps necessary to grow desired materials. Prior to executing a recipe, the automatic control program will conduct a system check to ensure that the furnace and butterfly valve are communicating with the program, the gases required for the recipe are available, the system is sealed, and the vacuum pump is operating. The system check can be followed by selecting the *'system check'* tab in the top left corner of program. If any of these checks fail, the CVD system is re-pressurized to atmospheric pressure with argon and the program is stopped so that the fail can be addressed. Once the system check is successful, the program will begin to run the recipe and the user can monitor the progress by selecting the *'run recipe'* tab to monitor the progress. A view of the automated control graphical user interface is shown in Fig 5. The right side of the interface is similar to the manual control program. However, the left side projects a table of the current stage and the corresponding set points for the selected recipe rather than user inputs. The graphical user interface also has a counter that records the time elapsed within the recipe as well as an estimated time of completion for the process. The user can also ensure that a log file is being recorded during the recipe by noting the blinking green box, right above the log file path in the lower left corner of the interface. The graphical user interfaces were designed to make the critical deposition conditions readily available and easily found during a process run.

**Table 1. Stages, or process steps, within the Quick.csv recipe.**

| Stage | 0 | 1 | 2 | 3 | 4 |
|---|---|---|---|---|---|
| Pressure Setpoint (Torr) | 680 | 500 | 300 | 0 | 680 |
| Temperature Setpoint (°C) | 25 | 200 | 500 | 500 | 25 |
| Temperature Ramp Rate (°C/min, 30°C/min max) | | | 30 | | |
| Ar flow (sccm) | 0 | 600 | 0 | 0 | 500 |
| $CH_4$ flow (sccm) | 0 | 0 | 400 | 0 | 0 |
| $H_2$ flow (sccm) | 0 | 0 | 0 | 200 | 0 |
| $C_2H_4$ flow (sccm) | 0 | 0 | 0 | 0 | 0 |
| Dwell time (min) | | | | 10 | |
| Stage End Condition | Start | Temp | Ramp | Time | End |

**Fig 5. Screen capture of the CVD system automated control graphical user interface showing recipe progress.**

# Results

Monolayer tungsten disulfide ($WS_2$) and $WS_2$/ graphene heterostructures were made using the CVD system built as described. For the heterostructures, graphene film was grown in the CVD chamber on copper foil (Alfa Aesar, 0.5 mm thick, 99.99%) and transferred to single-side polished sapphire (C plane) substrates (see complete preparation in S2 Appendix- Growth Details). Prior to $WS_2$ growth, the graphene film was annealed at 400 °C for 30 min under Ar/ $H_2$ flow to remove any residual PMMA [24] left over from the transfer process. The annealed graphene sample was then placed face down in a quartz boat and held 7.5 mm above tungsten oxide ($WO_{2.9}$)(Alfa Aesar) powder, which was placed in the center of the tube furnace. Upstream in a separate quartz boat, sulfur pieces of random sizes (Puratronic 99.9%, Alfa Aesar) were placed approximately 200 mm from the quartz boat containing the graphene sample, which was centered within the uniform hot zone of the furnace. The temperature of the sulfur was held at approximately 200 °C during the growth process, allowing the sulfur to vaporize. The atmospheric pressure CVD growth for WS$_2$ took place at 850 °C, with argon flowing at 70 sccm and hydrogen flowing at 15 sccm for 15 minutes. Similarly, we have also grown WS$_2$ on single-side polished sapphire (C plane) substrates using the same process set up and recipe, omitting the graphene steps. The growth recipe for WS$_2$ for use in the open-source automated CVD system is outlined in the supporting material (S2 Appendix- growth details).

## Tungsten Disulfide/ Graphene Heterostructures

The CVD growth of WS$_2$ on graphene/ sapphire substrates resulted in well-defined equilateral triangles, which is characteristic of single crystal domains [25] as shown in the optical micrograph (Zeiss Axio Imager.M2m) within Fig 6A. Raman spectra were collected (Horiba LabRAM) on graphene and WS$_2$ as shown in Fig 6B and 6C, giving an indication of the structural properties for the sample. Using an excitation source with a wavelength of 532 nm, the sample exhibits a strong peak at 1582 cm$^{-1}$, which corresponds to the G band, while the peak at 2700 cm$^{-1}$ is representative of the 2D band, as expected for graphene [26]. Near the edge of the graphene we also see a small peak at 1350 cm$^{-1}$ due to the disorder- induced D band. The presence of monolayer WS$_2$ is also validated through Raman spectroscopy using a 532 nm excitation wavelength as shown in Fig 6C. The spectra in Fig 6C include first order modes at the Brillouin zone center, $E_{2g}^1$, at 356 cm$^{-1}$ and A$_{1g}$ at 418 cm$^{-1}$, typical of monolayer WS$_2$ [27], [28].

**Fig 6. WS$_2$/ graphene heterostructures grown in the open-source CVD system.**
(A) Optical micrograph of 2D heterostructures showing WS$_2$ triangles. (B) Raman spectra of graphene. (C) Raman spectra of WS$_2$.

## Monolayer WS$_2$

Additionally, we demonstrate that controlled growth to a monolayer is possible with this open-source automated CVD system. First, the optical micrograph in Fig 7A confirms growth of $WS_2$ on sapphire in the characteristic triangular domains [25]. The thickness of as-grown $WS_2$ on sapphire was measured by atomic force microscopy (AFM). The height profile in the AFM image (Fig 7B) indicates a height difference of 0.7 nm as evidence of monolayer growth [25], [29].

**Fig 7. Synthesis of monolayer $WS_2$ on sapphire grown with the open-source CVD system.** (A) Optical micrograph showing $WS_2$ triangle domains. (B) AFM characterization of $WS_2$ on sapphire showing the height profile. (Bruker Nano Dimension FastScan) The AFM color scale represents the sample height with a range of 0- 145 nm.

## Discussion

The benefits of an open-source automated CVD system include: 1) a lower time investment for each growth cycle, 2) increased repeatability through the use of recipes, and 3) the ability to audit past growths since the automated program will record each growth cycle. Additional benefits of this system include the feasibility for a high level of customization, the ability to modify and update the system as needed, and a significant gain of experience in vacuum systems and system automation for those conducting the build. The open-source CVD system is also much cheaper than purchasing a commercial CVD, costing about $30,000 USD for parts as compared with $95,000 USD for a (low end) commercial turn-key system (quote obtained in 2015). Once you factor in the additional savings of time and labor, one can anticipate a savings exceeding half of the cost for a commercial system. The software for the system was also designed to be simple and easy to use, allowing for growths with minimal tuning of process variables.

## Conclusions

In summary, an open-source CVD system has been designed, sourced, built and programmed. This CVD system is the first open-source design that has shown comparable capabilities to commercially manufactured systems. This system is capable of both atmospheric pressure and low-pressure growths, temperatures up to 1100 ℃, and substrates up to ~20 mm in width. Additionally, with the detailed parameter and recipe logging, past growths can quickly be audited for possible issues or repeated, allowing for quick system troubleshooting and high process repeatability. With the system described here, we have grown 2D materials such as graphene, $WS_2$, as well as graphene/ $WS_2$ vertical heterostructures, with exceptional quality.

As open-source scientific hardware continues to increase, it is conceivable that the rate of scientific discovery could increase as researchers are freed to spend more time and effort conducting experiments, and less time building or troubleshooting a custom-built system. An open-source CVD system such as this one will help to propel 2D nanomaterials research forward by lowering financial barriers and allowing researchers with smaller budgets to join this emerging area of research. By sharing this design as an open-source, we hope to put the opportunity of research into many more labs, with a fairly reasonable startup cost thus accelerating the development and understanding of 2D materials.

## Acknowledgements

This work was supported through startup funds provided by the Micron School of Materials Science and Engineering at Boise State University and a gift from the Micron Foundation. D.B. acknowledges additional support through fellowships provided by the Idaho Space Grant Consortium and the Nuclear Regulatory Commission. The authors would also like to acknowledge Corey Efaw and Mike Hurley of the Applied Electrochemistry and Corrosion research lab for their assistance with the electrochemical transfer of graphene films. Atomic force micrographs were obtained through the Boise State University Surface Science Laboratory.

## Supporting information

**S1 Appendix. Bill of materials for CVD system divided by subsystem.**

**S2 Appendix. Construction notes.**

**S3 Appendix. Growth details.**

**S1 Folder Programs.** Folder contains all of the software needed to automate the CVD system.

**S2 Folder Support Drawings.** Folder contains construction drawings.

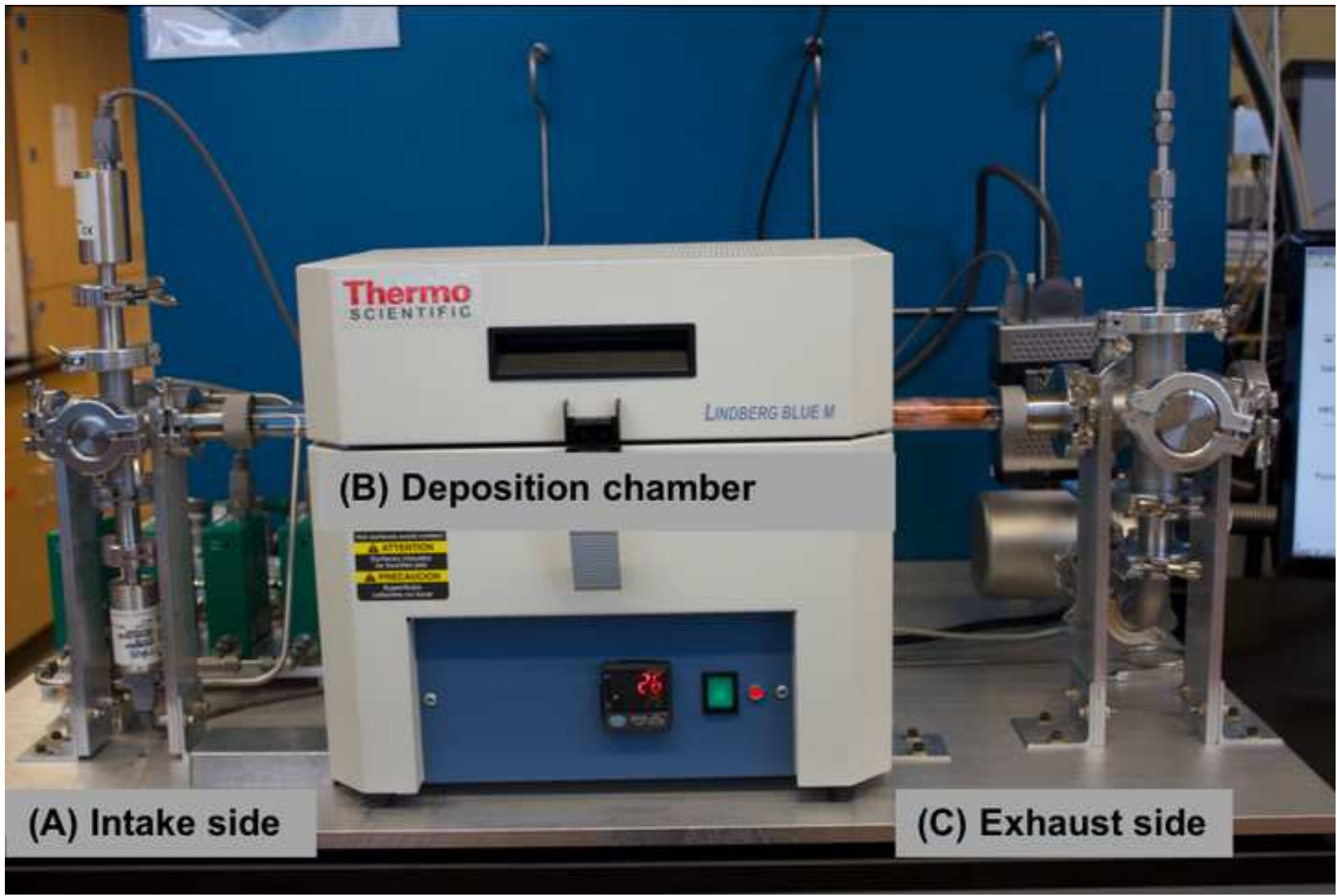

(A) Intake side  (B) Deposition chamber  (C) Exhaust side

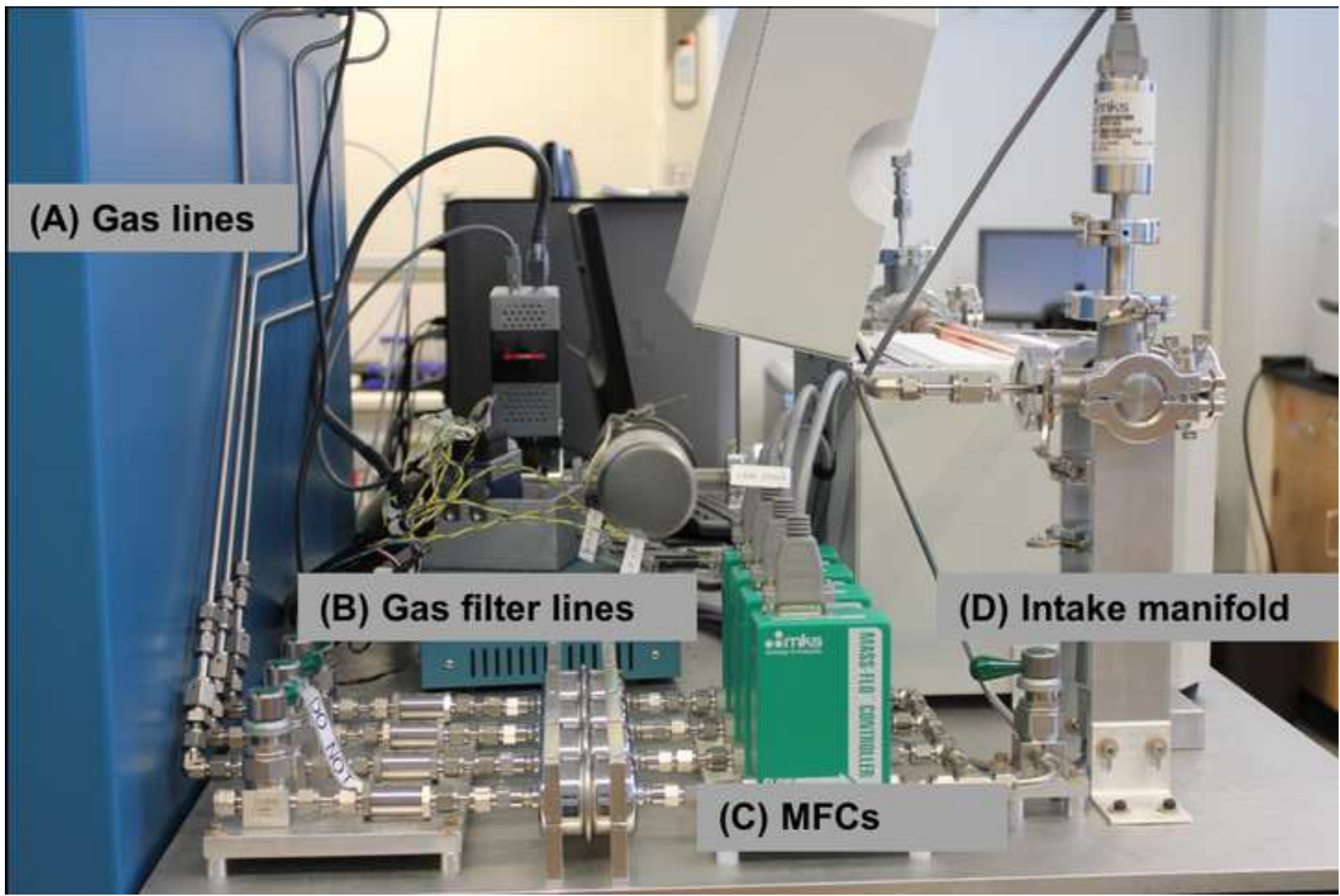

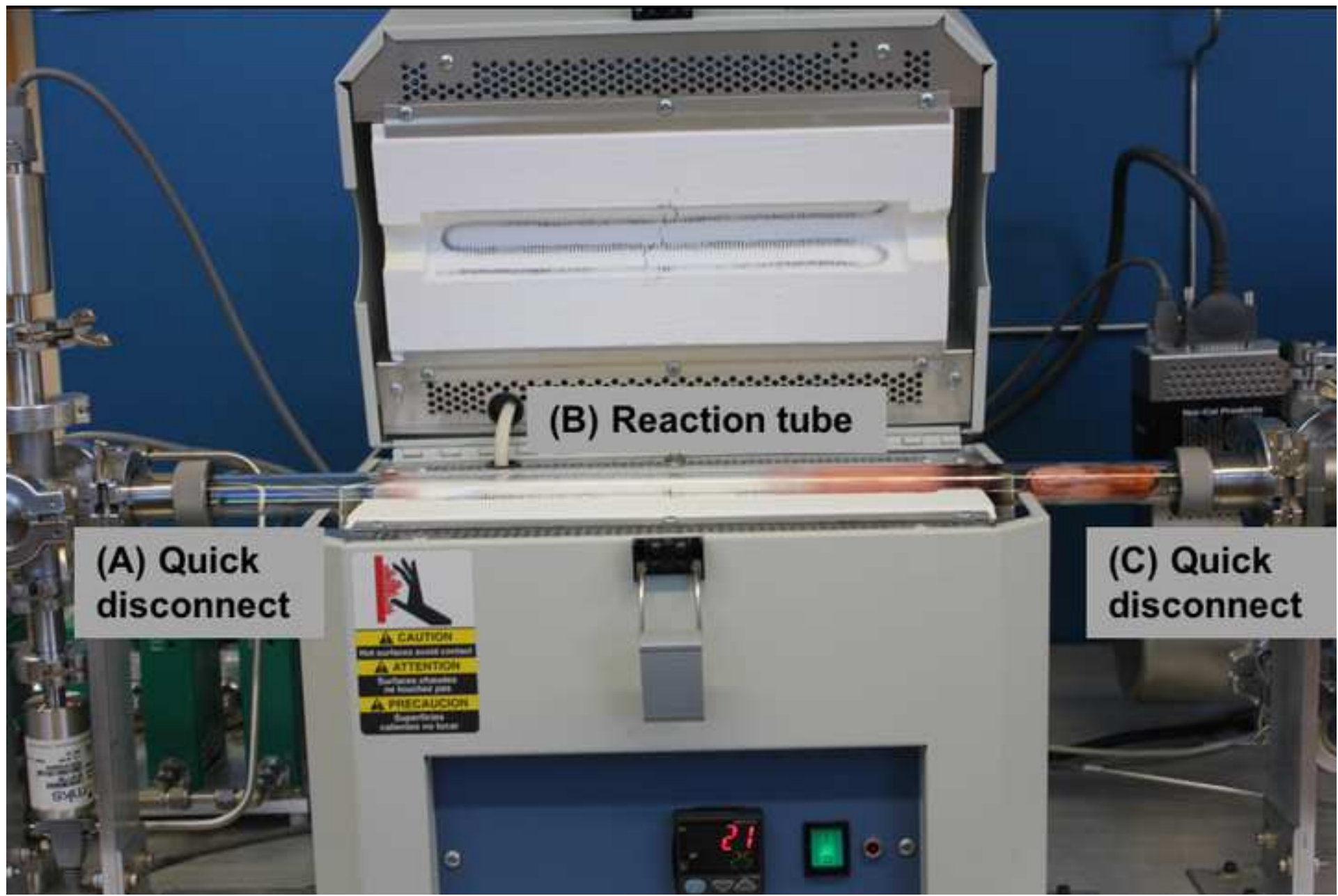

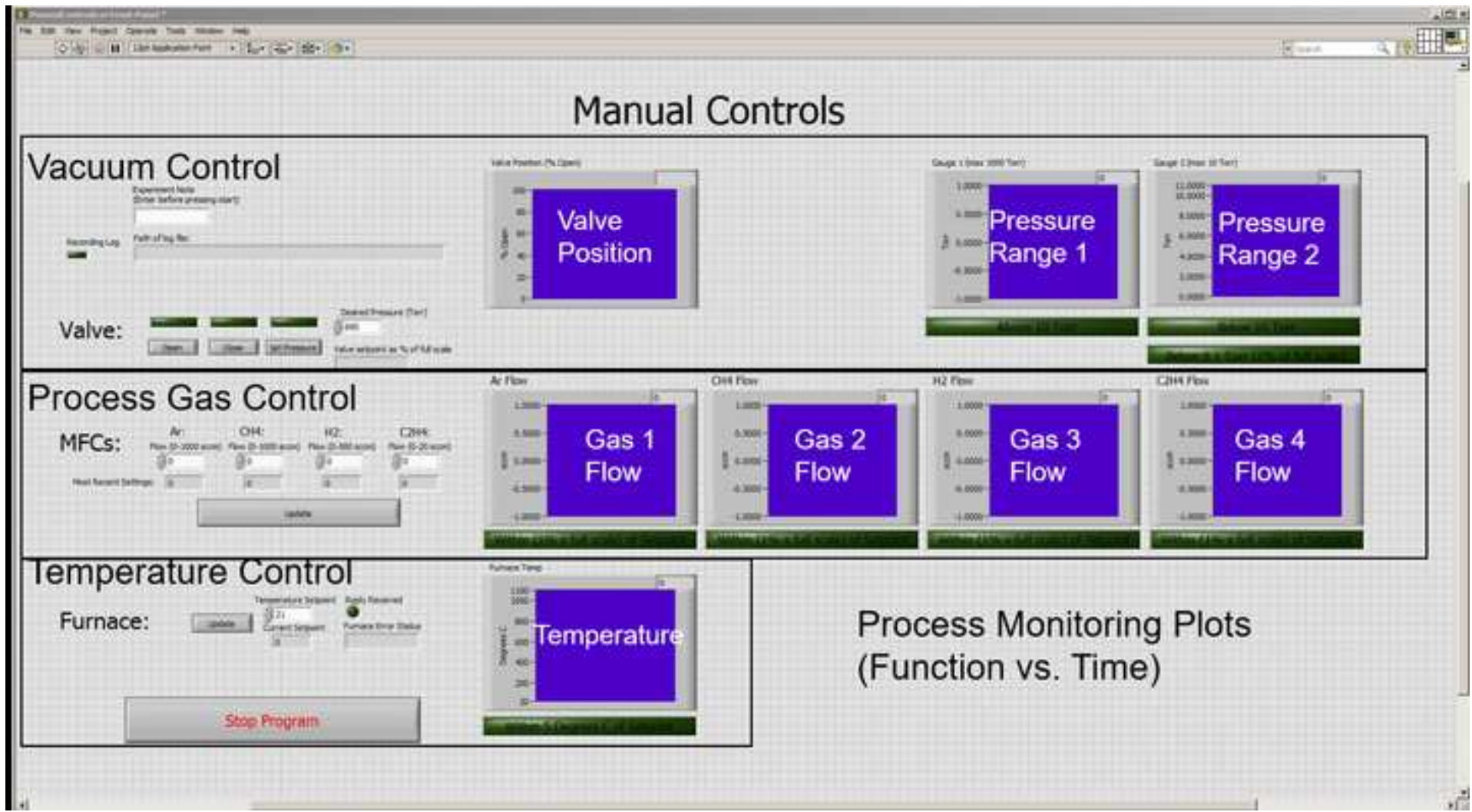

Figure                                                          Click here to download Figure Fig5.tif

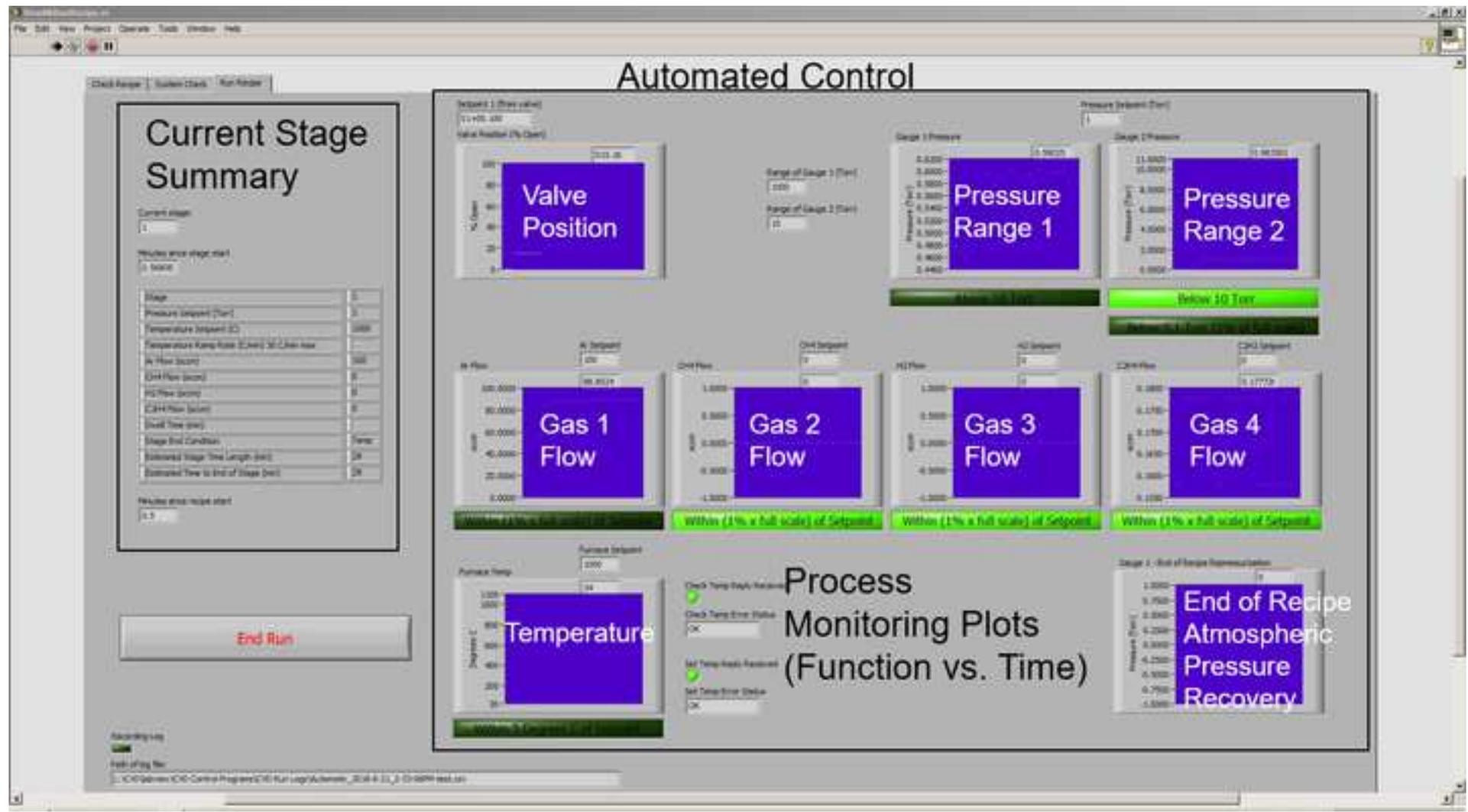

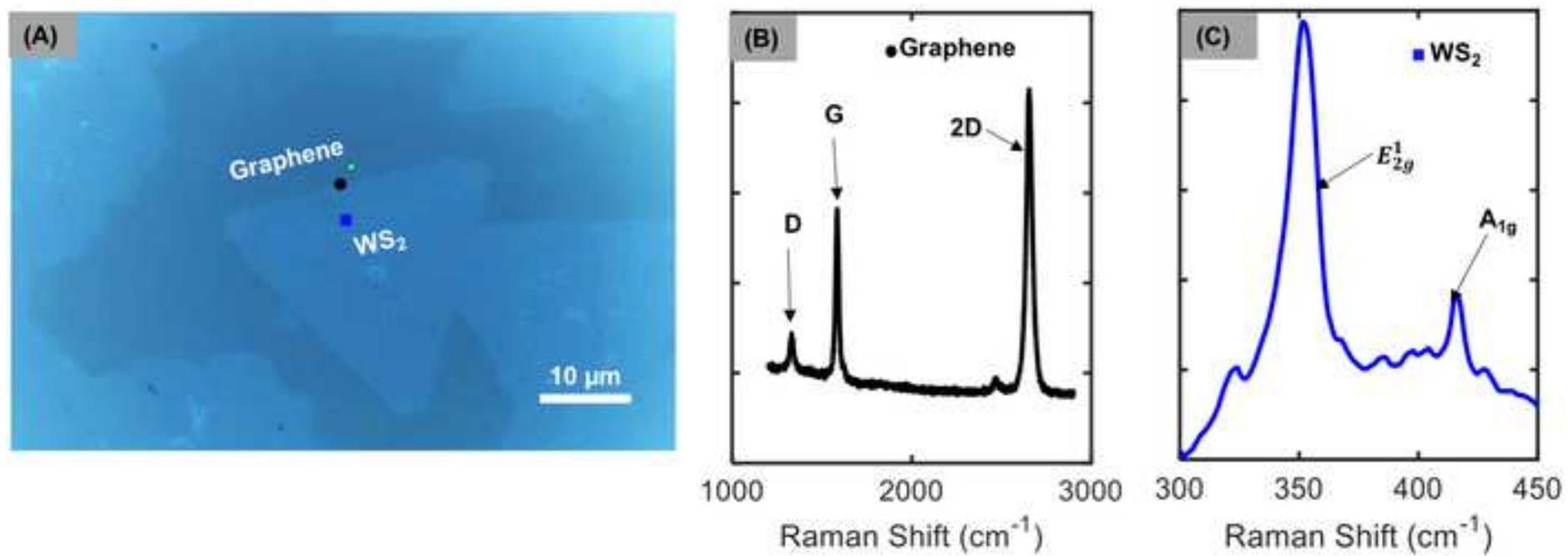

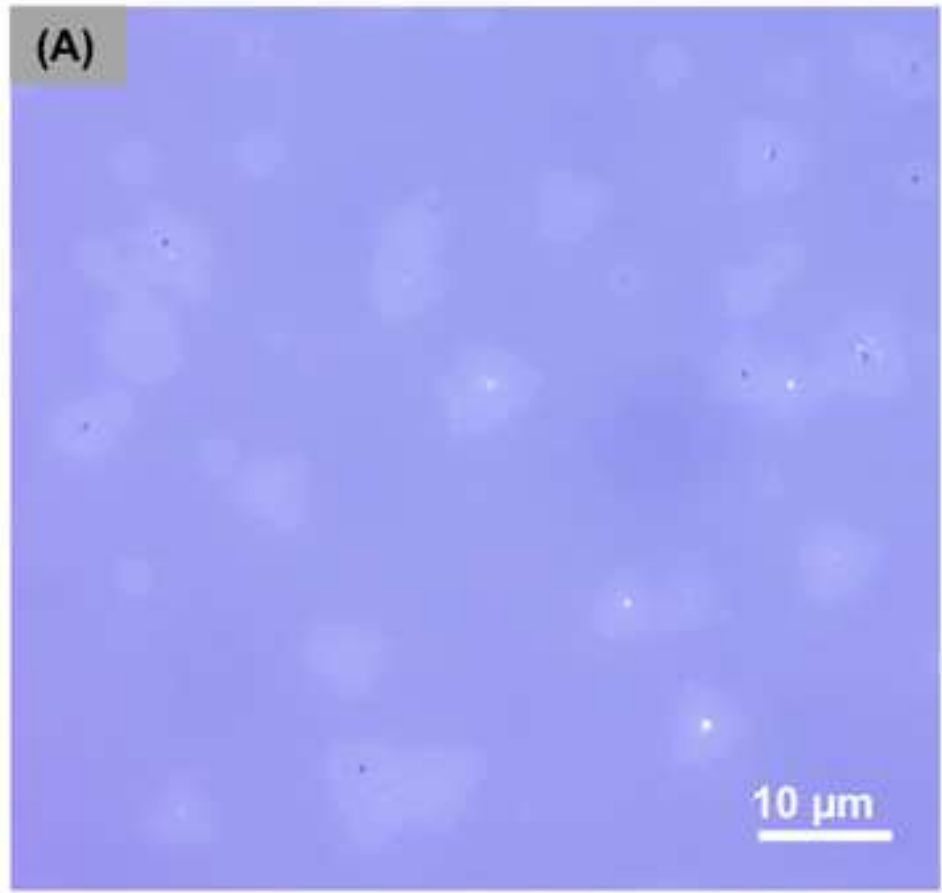
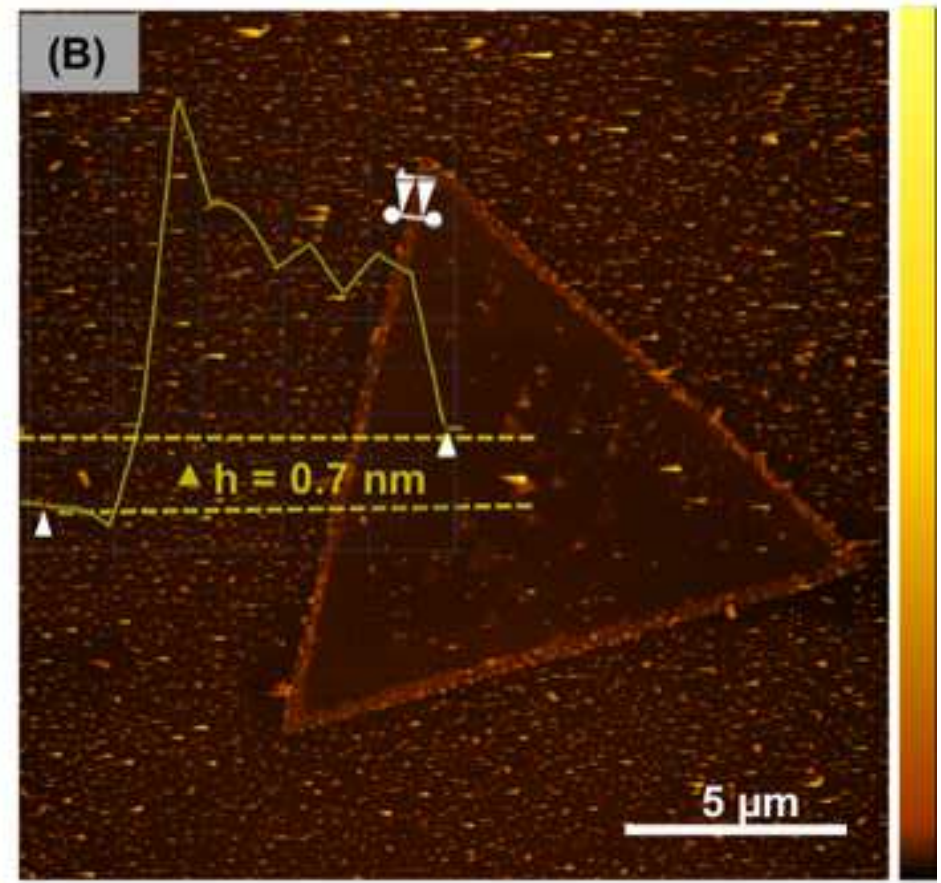

Supporting Information

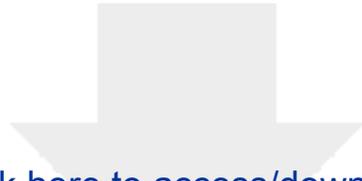

Click here to access/download
**Supporting Information**
S1 Appendix Bill of materials for CVD system.xls

Supporting Information

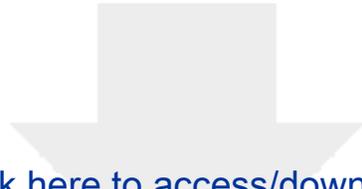
Click here to access/download
**Supporting Information**
S2 Appendix Construction Notes.pdf
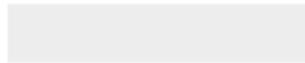
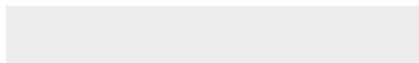

Supporting Information

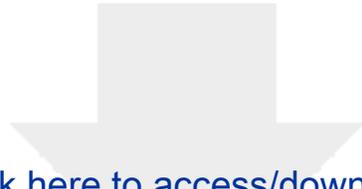
Click here to access/download
**Supporting Information**
S3 Appendix Growth Details.pdf
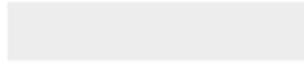
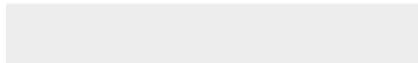

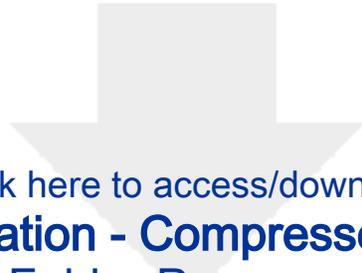

Click here to access/download
**Supporting Information - Compressed/ZIP File Archive**
S1 Folder Programs.zip

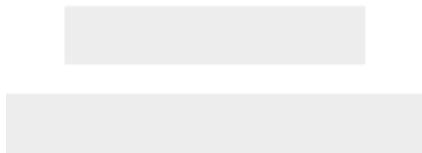

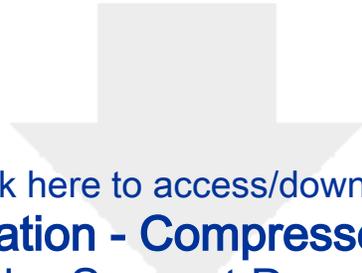

Click here to access/download
**Supporting Information - Compressed/ZIP File Archive**
S2 Folder Support Drawings.zip

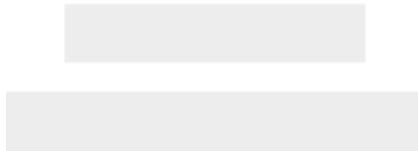